# Deep Uncertainty Surrounding Coastal Flood Risk Projections: A Case Study for New Orleans

**Tony E. Wong**[1] **and Klaus Keller**[1,2,3]

[1]Earth and Environmental Systems Institute, Pennsylvania State University, University Park, PA 16802, USA.

[2]Department of Geosciences, Pennsylvania State University, University Park, PA 16802, USA.

[3]Department of Engineering and Public Policy, Carnegie Mellon University, Pittsburgh, PA 15289, USA.

Corresponding author: Tony Wong (twong@psu.edu)

**Key Points:**

- We characterize key deep uncertainties surrounding flood risk projections for a levee ring in New Orleans using 18 probabilistic scenarios

- The levee system alone likely provides flood protection between the 100- and 500-year return period

- Uncertainty in the storm surge distribution shape parameter is the primary driver of flood risk variability

Abstract

Future sea-level rise drives severe risks for many coastal communities. Strategies to manage these risks hinge on a sound characterization of the uncertainties. For example, recent studies suggest that large fractions of the Antarctic ice sheet (AIS) may rapidly disintegrate in response to rising global temperatures, leading to potentially several meters of sea-level rise during the next few centuries. It is deeply uncertain, for example, whether such an AIS disintegration will be triggered, how much this would increase sea-level rise, whether extreme storm surges intensify in a warming climate, or which emissions pathway future societies will choose. Here, we assess the impacts of these deep uncertainties on projected flooding probabilities for a levee ring in New Orleans, Louisiana. We use 18 scenarios, presenting probabilistic projections within each one, to sample key deeply uncertain future projections of sea-level rise, radiative forcing pathways, storm surge characterization, and contributions from rapid AIS mass loss. The implications of these deep uncertainties for projected flood risk are thus characterized by a set of 18 probability distribution functions. We use a global sensitivity analysis to assess which mechanisms contribute to uncertainty in projected flood risk over the course of a 50-year design life. In line with previous work, we find that the uncertain storm surge drives the most substantial risk, followed by general AIS dynamics, in our simple model for future flood risk for New Orleans.

# 1 Introduction

Sea-level rise poses nontrivial risks to coastal communities [*Nicholls and Cazenave*, 2010; *Hallegatte et al.*, 2013; *Hinkel et al.*, 2014, 2015]. The performance of strategies to manage these risks relies on sound projections of future sea levels and storm surges, including the inherent uncertainties [cf. *Hinkel et al.*, 2015].

In a comparison of projected future flood losses in 136 coastal cities across the world, recent work has found that New Orleans (Louisiana, United States of America) ranks second in average annual loss (as percentage of the city's gross domestic product), even with adaptation to maintain present flood risk [*Hallegatte et al.*, 2013]. Another recent study found substantially higher estimates for the average annual losses faced by New Orleans [*Abadie et al.*, 2017], where the discrepancy between the two studies is driven by the higher projected sea-level rise used in the latter study [*Kopp et al.*, 2014]. The United States Army Corps of Engineers (USACE) has stated its objective to develop coastal defense strategies (including for New Orleans) that perform well across a range of plausible climate change scenarios [*Moritz et al.*, 2015]. However, recent scientific findings suggest that future flood risks may be higher and more uncertain than previously estimated [e.g., *Grinsted et al.*, 2013; *DeConto and Pollard*, 2016; *Abadie et al.*, 2017; *Bakker et al.*, 2017; *Wong et al.*, 2017b]. Here, we assess the expected performance of the flood protection system in New Orleans under a set of plausible future climate scenarios that account for key new scientific findings and sample key deep uncertainties. We caution that our intent is to attribute and characterize the impacts of these deep uncertainties, and are not meant to directly inform on-the-ground decisions.

Deep uncertainty refers to the situation in which experts cannot agree on the set of possible outcomes, the consequences of those outcomes, and/or the associated probabilities [e.g., *Langlois and Cosgel*, 1993; *Walker et al.*, 2013]. In these cases, the use of a range of plausible future scenarios or multiple probability density functions (pdfs) can provide useful inputs for the design of risk management strategies [e.g., *Schoemaker*, 1993; *Bryant and Lempert*, 2010; *Lempert et al.*, 2012]. Such a representation of uncertainties can be critical to inform decision-making, for example in the context of coastal risk management [*De Winter et al.*, 2017]. The presentation of multiple scenario pdfs enables the identification of scenarios to which particular strategies may be vulnerable, and for decision-makers to determine the appropriate protective action in the face of these vulnerabilities [see, for example, *Lempert et al.*, 2012]. Furthermore, using multiple pdfs can provide a traceable combination of information from diverse sources and containing differing levels of uncertainty [*Lempert et al.*, 2012] and characterizes the impacts of deep uncertainty on possible future outcomes [*Oppenheimer and Alley*, 2016].

The deeply uncertain future radiative forcing is often represented in the form of Representative Concentration Pathways (RCP) scenarios [*Meinshausen et al.*, 2011]. These scenarios may be used to link radiative forcing to the consequent changes in local sea levels, and finally to their implications for coastal defense strategies [*Jackson and Jevrejeva*, 2016; *Wong et al.*, 2017b]. Uncertainty in future storm surge severity also drives substantial risks to coastal communities [*Resio et al.*, 2013]. Previous approaches to project future storm surges have included physical modeling [e.g., *Orton et al.*, 2016], statistical modeling [e.g., *Grinsted et al.*, 2013], and scenarios [e.g., *Lempert et al.*, 2012; *Johnson et al.*, 2013]. Here, we expand upon and combine these approaches by providing probabilistic scenarios of future storm surge level. Potential fast Antarctic ice sheet disintegration via cliff instability and hydrofracturing in response to rising global temperatures is a critical deep uncertainty, driving potentially large changes in sea-level rise this century [*DeConto and Pollard*, 2016; *Oppenheimer and Alley*, 2016; *Bakker et al.*, 2017; *Ruckert et al.*, 2017; *Wong et al.*, 2017b]. Recent work projecting this fast dynamical Antarctic contribution to sea level used radiative forcing scenarios [*Meinshausen et al.*, 2011] as well as multiple prior distributions on key model parameters to provide probabilistic ranges within each scenario [*Wong et al.*, 2017b]. Other recent studies employed statistical approaches based on the projections of *DeConto and Pollard* [2016] to construct probabilistic projections of the rapid Antarctic mass loss [*Kopp et al.*, 2017; *Le Bars et al.*, 2017]. Here, we employ the mechanistically-motivated emulator of *Wong et al.* [2017b] to construct our projections of future Antarctic contributions to sea level. This model choice is motivated by (i) our intention to explicitly represent deeply uncertain geophysical mechanisms (e.g., the triggering of Antarctic fast ice loss) and connect these to local flood risk, and (ii) a successful hindcast test [*Wong et al.*, 2017b]. We caution, however, that passing a hindcast test [see, for example, *DeConto and Pollard*, 2016; *Wong et al.*, 2017b] is, of course, no guarantee that a model will capture future dynamics, especially when the physical system in question is as deeply uncertain and nonlinear as the Antarctic ice sheet.

There are, of course, other deep uncertainties affecting flood risk that could be taken into account in future work. These include potential rapid ice melt/disintegration from the Greenland ice sheet, more complex basin-scale Antarctic dynamics [*Ritz et al.*, 2015], as well as potential feedbacks and correlations among the deeply uncertain factors considered here (e.g., between RCP scenario and storm surge non-stationarity). The fragility of the flood protection system may also be considered a deeply uncertain factor, for example, to bound the probability of failure [*Johnson et al.*, 2013]. In this study, we consider only the overtopping levee failure mechanism. This provides a reasonable estimate of a lower bound on the failure probability. This serves to simplify the analysis and leaves the treatment of levee reach-scale failure properties to more detailed analyses [e.g, *Fischbach et al.*, 2012].

Here, we provide probabilistic projections of flood risk for the city of New Orleans, conditioned on the deeply uncertain scenarios detailed above. Specifically, we consider multiple future scenarios spanning radiative forcing, fast Antarctic ice sheet disintegration, and characterization of future storm surges. We expand upon previous work by providing a probabilistic accounting of both stationary and non-stationary storm surge characteristics, as well as putting recent scientific findings – the potential fast Antarctic ice sheet dynamic mass loss contributions to sea-level rise – into a local coastal protection context. We use a global sensitivity analysis [*Sobol'*, 2001] to identify the key modeling uncertainties that drive uncertainty in future flood risk. The scope of this work is limited to the characterization and attribution of the impacts of deeply uncertain geophysical drivers on uncertainty in projected flood risk. The purpose of this work is (i) to provide a menu of plausible probabilistic scenarios of future flood risk for New Orleans, (ii) highlight primary drivers of flood risk, and (iii) to incorporate into the flood risk assessment the newly available probabilistic projections of the fast Antarctic ice sheet contribution to sea levels.

## 2 Methods

2.1 Sea-level rise

We simulate global mean surface temperature, ocean heat uptake, and global mean sea level and its contributions from the Antarctic ice sheet, Greenland ice sheet, thermal expansion, glaciers and ice caps, and land water storage using the Building Blocks for Relevant Ice and Climate Knowledge (BRICK) model v0.2 [*Wong et al.*, 2017a] with a modified Antarctic ice sheet (AIS) fast dynamics emulator [*Wong et al.*, 2017b]. BRICK is an open source platform of semi-empirical models for the major contributions to global mean sea level. We calibrate the model preferentially to observational records as opposed to simulation output from more complex models. The model set-up, calibration to paleoclimate and instrumental data, and simulations proceed as described in *Wong et al.* [2017b]. We run the model from 1850 to 2065. We use 2065 as the time horizon for analysis in light of the 50-year planning horizon implemented in the Louisiana Coastal Protection Master Plan [p. 48, *Coastal Protection and Restoration Authority of Louisiana*, 2017]. Sea-level rise is presented relative to sea level in 2015. We use historical radiative forcing (described by *Urban and Keller* [2010]) and observational data between 1850 and 2009 to calibrate the model [*Wong et al.*, 2017a]. The model is then run from 1850 to 2065 under the RCP2.6, 4.5, and 8.5 radiative forcing scenarios [*Meinshausen et al.*, 2011]. Thus, there are three scenario cases for radiative forcing. Local scaling factors are used to estimate the local mean sea-level rise in New Orleans, from the simulated contributions to global sea level from glaciers and ice caps, the Greenland ice sheet, the Antarctic ice sheet, thermal expansion, and land water storage [*Slangen et al.*, 2014; *Wong et al.*, 2017a]. We sample the effects of land subsidence on local sea level around a recent estimate [*Dixon et al.*, 2006] (see the discussion in Section 2.3). All sea levels presented are New Orleans local sea level.

Recent work has shown that ice cliff instability and hydrofracturing may lead to substantial and rapid contributions from the Antarctic ice sheet to sea-level rise [*DeConto and Pollard*, 2016]. We employ the following emulator for the contribution of these fast Antarctic ice sheet dynamics ("AIS fast dynamics") to global mean sea level:

$$\frac{dV}{dt} = \begin{cases} -\lambda, & T > T_{crit} \\ 0, & T \leq T_{crit} \end{cases}, \tag{1}$$

where $\lambda$ is an uncertain rate of disintegration (mm y$^{-1}$), $T_{crit}$ is the threshold temperature at which fast dynamical disintegration is triggered (°C), and $T$ is the mean Antarctic surface temperature, reduced to sea level [*Diaz and Keller*, 2016; *Wong et al.*, 2017b].

$\lambda$ and $T_{crit}$, along with 39 other BRICK model parameters, are jointly estimated within a Bayesian calibration framework [*Wong et al.*, 2017a]. We include three scenarios for the potential AIS fast dynamics: "*gamma*", "*uniform*", and "*none*". In the "*gamma*" and "*uniform*" scenarios, the prior distributions assumed for the fast dynamics parameters $\lambda$ and $T_{crit}$ are assumed to be gamma or uniform distributions, respectively [*Wong et al.*, 2017b]. The effect of the gamma prior is to put more weight on central values of these parameters, with ranges informed by the literature [*DeConto and Pollard*, 2016]. Each of these two sets of prior distributions yields an ensemble of calibrated projections of future sea level, including the additional sea-level contributions from the AIS fast dynamics. These ensembles each contain

12,586 states-of-the-world (SOW), and are simulated under each of the three radiative forcing scenarios.

The ensemble of simulations for the third AIS fast dynamics scenario, "*none*", is drawn from the *gamma* and *uniform* ensembles (6,293 simulations each), neglecting the contributions of the AIS fast dynamics to sea level. The three AIS fast dynamics scenarios span a range of possible outcomes for the fast dynamical AIS contributions to sea level this century – *none*, AIS disintegration does not occur; *uniform* – we possess enough information to bound the values of the parameters driving disintegration; *gamma* – we possess a high degree of confidence in our prior beliefs regarding the disintegration parameterization. We run each of these three AIS fast dynamics scenarios under each of the three radiative forcing scenarios. This yields a total of nine sea-level rise scenarios.

2.2 Storm surge

We use tide gauge data from Grand Isle, Louisiana and a Bayesian calibration approach to fit an ensemble of stationary generalized extreme value (GEV) distributions for the storm surge level [*NOAA*, 2017]. The hourly tide gauge data span the period from November 1980 to December 2016. First, we subtract annual means from the tide gauge record. Second, we calculate annual block maxima from the detrended record. Third, we calculate a maximum likelihood estimate for the GEV distribution parameters, fit to the tide gauge data (see Supplemental Text S1). Fourth, we use the resulting estimate as the starting point for two 500,000-iteration Markov chain Monte Carlo simulations of the GEV parameters (see Supplemental Text S1). Finally, we discard the first 100,000 iterations of each chain for "burn-in" (see Supplement) and draw an ensemble of 12,586 samples from the combined sample of 800,000 remaining chain members. The ensemble size is dictated by the size of the BRICK sea-level rise ensemble of simulations. The extreme value analysis described here is done using the *extRemes* package in the R Programming Language [*Gilleland et al.*, 2013; *R Core Team*, 2016]. These 12,586 sets of GEV parameters are used to characterize the stationary (i.e., not time-varying) distribution of storm surge level in New Orleans for each of the radiative forcing/AIS fast dynamics scenario combinations. Thus, there are nine scenarios where the distribution of storm surge levels is assumed to remain stationary, characterized by their present values but accounting for uncertainty through the Markov chain Monte Carlo sampling.

We follow an approach outlined by the United States Army Corps of Engineers [*USACE*, 2012] to extrapolate the potential for non-stationary storm surge behavior from the stationary storm surge characterization and changes in future sea levels. Specifically, we project the increase in surge level as the product of the increase in sea level and an uncertain "surge factor", $C_{surge}$, which we sample uniformly between 1.5 and 2 [Table 1.2, *USACE*, 2012]. The range of values for $C_{surge}$ is based on a set of Advanced Circulation Model (ADCIRC) simulations, which were conducted to assess the effects of the increasing sea levels on surge heights [*USACE*, 2012]. We calculate the non-stationary storm surge level as the sum of the storm surge level from the stationary GEV distribution and the increase in surge levels. Thus, a distribution for the increase in storm surge level provides a non-stationary storm surge characterization. This gives rise to nine scenarios (each of the radiative forcing and AIS fast dynamics combinations) with non-stationary storm surge, and a total of 18 scenarios.

Previous work has projected changes in future storm surge levels by fitting a non-stationary GEV distribution to a sizeable tide gauge record [*Grinsted et al.*, 2013]. Alas, the Grand Isle data set contains only 36 full years of data. The discussion in *Grinsted et al.* [2013] and preliminary experiments conducted here suggest that such a short record is not sufficient to constrain a fully non-stationary GEV model with high precision. Indeed, the non-stationary storm surge is comprised of two components, one driven by natural variability and the other driven by long-term climate changes. The former is likely poorly sampled in such a short tide gauge record. We hence follow the alternative approach described above [*USACE*, 2012]. This approach misses potentially important processes, such as the reported influence of elevated temperatures on storm surge intensity [e.g., *Grinsted et al.*, 2013], or changes in the Atlantic Meridional Overturning Circulation and upper ocean heat uptake [e.g., *Little et al.*, 2015]. Our simple approach does link sea-level changes (driven by changing global temperatures) to the simulated storm surge. In this approach, sea level can be viewed as a proxy for temperature, in driving increased storm surge intensity (though increased temperature does not necessarily always lead to increased surge intensity), however this approach still misses the potential changes in storm surge driven by mechanisms other than temperature.

2.3 Coastal flood risk

We focus on the north-central Metropolitan New Orleans levee ring. This levee system contains the East Jefferson and Orleans Metropolitan areas and has been the subject of previous studies [e.g., *Jonkman et al.*, 2009; *Wong et al.*, 2017b]. We assume a levee ring average height of 4.88 meters (16 feet) and an initial levee structure base elevation relative to local sea level of -1.22 meters (-4 feet) [*USACE*, 2014]. That the base of the flood protection structure is negative indicates the effects of land subsidence. We sample uncertainty in future land subsidence by drawing 12,586 samples from a log-normal distribution with mean 5.6 mm $y^{-1}$ and standard deviation 2.5 mm $y^{-1}$ [*Dixon et al.*, 2006], where the size of the ensemble is determined by the size of the BRICK sea-level rise ensemble of simulations.

Our method produces 18 probabilistic scenarios that include distributions of (i) sea-level rise, (ii) land subsidence, (iii) stationary storm surge, and (iv) increase in surge level (for the non-stationary scenarios only). These scenarios are summarized in Table 1. We note that these scenarios do not cover the set of all possible futures, and that the probabilistic projections within each scenario are conditionally dependent on the deeply uncertain storm surge, AIS fast dynamics, and RCP forcing, which do not have probability distributions assigned to them. Thus, these 18 scenarios serve to *characterize* key aspects of these deep uncertainties as opposed to *quantify* them. The total sea plus storm surge level that is incident on the flood protection system is the sum of these individual components. We focus our attention on these distributions of flood risk in the year 2065.

We calculate flood probabilities for each year as the tail area of the local mean sea plus storm surge level distribution above the levee height (reduced for the initial subsidence), to yield the annual exceedance probability. We estimate the return period as the inverse of the average annual exceedance probability. For example, the Coastal Protection and Restoration Authority of Louisiana has target protection levels of 1/100 annual exceedance probability (100-year return period) for general projects and 1/500 annual exceedance probability (500-year return period) suggested for critical infrastructure such as hospitals and emergency response stations [p. 143,

*Coastal Protection and Restoration Authority of Louisiana*, 2017]. Note that the flood probability in 2065 is almost certainly larger than the average annual exceedance probability due to temperature-driven increases in sea level and storm surge intensities [*Grinsted et al.*, 2013; *Kopp et al.*, 2016]. Note further that the failure probabilities calculated here consider only the protection offered by the levee system; other adaptive and protective measures may increase the effective return period. Additionally, we only consider the overtopping failure mode at present. Levee failure modes such as piping and slope instability lead to an actual failure probability that is likely larger than the estimates presented here (see Discussion). Our analysis also does not account for the potentially sizable cumulative hazard arising from frequent minor flooding, or nuisance flooding [*Moftakhari et al.*, 2017]. As sea-level rise reduces the buffer between sea level and the flood stage, nuisance flooding is expected to become more frequent and more severe over the next few decades [*Ray and Foster*, 2016; *Vandenberg-Rodes et al.*, 2016], and will likely surpass the 30 days/year tipping point by mid-century [*Sweet and Park*, 2014]. This is, of course, an important avenue for continued study.

2.4 Sobol' sensitivity analysis

We employ an approach for global sensitivity analysis based on the method originally described by *Sobol'* [2001], and expanded by *Saltelli* [2002] and *Janon et al.* [2014]. These works describe the Sobol' sensitivity analysis method in detail; we hence provide a brief overview here. The approach decomposes the variance in a model response into components attributable to individual uncertain model parameters or forcings. We decompose the variance in the modeled mean annual exceedance probability (i.e., the flood risk) over the 2015-2065 period. We attribute the variance to uncertain endogenous model parameters (pertaining to land subsidence, storm surges, and changes in global temperature and the associated glacier, ice sheet, and ocean thermal responses), as well as to exogenous model forcings, including heightening of the levee system (by construction) and radiative forcing pathway (RCP scenario). We sample RCP scenario uniformly from [0, 1], where each of four intervals of width 0.25 correspond to the scenarios RCP2.6, 4.5, 6.0, and 8.5. Levee heightening is sampled uniformly from [-0.91 m, 0.91 m]. This range is chosen to reflect current plans to heighten much of this levee system and the fact that the levee height is not uniform around the levee ring [e.g., Table 3-8, *USACE*, 2014].

We calculate first-order Sobol' sensitivity indices (direct parameter/forcing influences on projected flood risk variance) and total sensitivity indices (the sum of the first- and all higher-order interaction indices for a particular term) for each parameter/forcing term, and the second-order sensitivity indices (the impacts on flood risk of interactions between two parameters/forcings) for each parameter/forcing pair. We display the results for these sensitivity indices in a radial convergence diagram [*Lima*, 2011]. In the radial convergence diagram, each parameter/forcing term is represented by a node around the perimeter of a circle, grouped by sub-model component. The color of the node indicates first-order (direct) influence on flood risk or total sensitivity (direct plus all indirect influences via parameter/forcing interactions). The size of each term's node corresponds to the magnitude of the flood risk sensitivity to that term – larger nodes indicate greater sensitivity. Bars connecting nodes indicate that interactions between two terms are connected to flood risk variability; thicker bars indicate greater second-order sensitivity of flood risk to interactions between that parameter/forcing pair. Note that the sum of all of the first-, second-, third- (and so on) indices necessarily equals 100%. The sum of the total sensitivity indices, however, exceeds 100% due to multiple counting of the higher-order interactions among the parameters and forcings. The Supporting Information provides a table containing the parameter and forcing term symbols and descriptions.

We construct an ensemble of 10,248,000 model realizations, using model parameters drawn from the distributions resulting from the calibration to paleoclimate and instrumental data (as described in *Wong et al.* [2017b]). We use bootstrap resampling to estimate confidence intervals for each sensitivity index. We select sample sizes to yield confidence intervals that are all <10% of the total sensitivity index for the leading index [*Butler et al.*, 2014], using 50,000 bootstrap replicates. We only report sensitivity indices above a threshold of 1% of the total variance.

## 3 Results

### 3.1 Probabilistic projections of future sea level and storm surge

We characterize the deeply uncertain future sea and storm surge levels using probabilistic projections within each of the 18 scenarios (Figure 1). The impacts of the AIS fast dynamics on future sea level are noticeable in RCP4.5 and 8.5 at the 1/500 level (Figure 1c). At the 1/100 level, the AIS fast dynamics drive substantial sea-level changes by 2065 under RCP8.5, (Figure 1a, c). This conclusion is in line with the results of *Wong et al.* [2017b], who find a time horizon of 2060 (ensemble median) for the onset of fast dynamical AIS disintegration under RCP8.5. These effects are dwarfed by the projected increases in storm surge levels, however (Figure 1b, d).

The tail events in non-stationary storm surge threat are much larger than those from sea-level rise alone (i.e., low-probability but high-impact events such as fast AIS disintegration). At the 1/100 level, the AIS fast dynamics drive considerable increases in sea level in RCP8.5 (Figure 1c). But the risks posed by non-stationary storm surge at the 1/100 level are much higher (Figure 1d). This gives rise to the structure of Figure 2: fast AIS disintegration occurs in the most severe SOW, but the driving force behind high flood risk is the storm surge severity. In particular, the non-stationary storm surge scenarios lead to the highest flood risks (Figure 2b).

### 3.2 Potential failure to meet flood protection standards

The 100-year flood protection standard (1/100 mean annual exceedance probability) [p. 143, *Coastal Protection and Restoration Authority of Louisiana*, 2017] is achieved by the levee system alone in the ensemble median in all considered scenarios (Figure 3, solid red vertical line). A sizeable number of SOW in each scenario extend below the 100-year protection standard, however. The left-most extent of each scenario box in Figure 3 is the 25%-quantile for that scenario's SOW. This implies that across all scenarios, the levee system alone does not meet the minimum standard of protection in roughly 25% of the SOW (the range across the scenarios is 21-28%).

The 500-year flood protection suggested for critical infrastructure (1/500 mean annual exceedance probability) [p. 143, *Coastal Protection and Restoration Authority of Louisiana*, 2017] is not met by the levee system alone in the ensemble median in any scenario (Figure 3, dashed yellow vertical line). In each of the three RCP8.5 scenarios combined with non-stationary storm surge (Figure 3, top three rows), about 65% of the SOW fail to meet this protection standard (the range across all scenarios is 61-68%). No SOW in any scenario meets the economically-efficient return level of 5,000 years, as found by *Jonkman et al.* [2009] (Figure 3, dot-dashed blue vertical line).

The USACE is planning to raise much of this levee system by about 0.91 meters (3 feet), in addition to other adaptive flood protection measures [*USACE*, 2014; *Moritz et al.*, 2015]. Even with the additional heightening, the levee system alone may still not be sufficient to attain required protection levels for critical infrastructure will likely fall sizably short of economic efficiency (see Supporting Information, Figure S1). The focus on the 50 year time horizon is motivated by the time horizon considered by the Coastal Protection and Restoration Authority of Louisiana [*Coastal Protection and Restoration Authority of Louisiana*, 2017]. This is relatively short compared to the time scale of the committed sea-level response. Recent work projecting future sea level contributions from fast Antarctic ice sheet dynamics confirms that the 50-year time horizon likely misses substantial sea-level contributions during the second half of this century [*DeConto and Pollard*, 2016; *Bakker et al.*, 2017; *Le Bars et al.*, 2017; *Wong et al.*, 2017b]. This underscores the need for adaptive strategies for coastal risk management [*Moritz et al.*, 2015]. Previous work has demonstrated that a "future without action" fails to protect the Greater New Orleans area to the 100-year level by 2061 [*Johnson et al.*, 2013]. These authors considered a larger potential flood zone and uncertainty in system fragility beyond only overtopping of the levee system considered in the present work, which accounts for the differing estimates of system reliability. We note that the high-risk upper tail of our sea-level projections under RCP8.5 with fast Antarctic dynamics (c.f. Figure 1c) fall roughly between their "less-optimistic" and "high sea-level rise" scenarios of 45 and 78 cm sea-level rise over the 50-year time horizon.

3.3 Sensitivity analysis

We find that the storm surge dominates the decomposition of variance in projected flood risk (Figure 4). The first-order sensitivity index for the shape parameter $\xi$ of the storm surge GEV distribution accounts for 77% of the total variance in flood risk. This result is expected because the shape parameter determines the heaviness of the high-risk upper tail of the GEV distribution. This result is also in agreement with other recent assessments regarding the dominance of storm surge in flood risk uncertainty [*Le Cozannet et al.*, 2015; *Oddo et al.*, 2017]. It is also difficult to constrain the shape parameter with observational data, because extremal data are by definition rare. Other statistical approaches that make use of more data than the annual block maximum approach employed here may be of use to further constrain this parameter (e.g., monthly blocks or a Poisson process/generalized Pareto distribution approach [e.g., *Tebaldi et al.*, 2012]). The first-order index for levee height ("build") accounts for another 8% of the flood risk variance. Perhaps not surprisingly, the total sensitivity index for "build" is 15%, demonstrating substantial interaction between levee heightening and other drivers of flood risk variability. Specifically, the second-order interaction index between "build" and the GEV shape parameter is 5% of the total variance in flood risk. The only other statistically significant first-order sensitivity index is the scale parameter of the storm surge GEV distribution, at 6% of the total flood risk variance.

The modeled global mean climate contributes substantially to the variance in flood risk via climate sensitivity ($S$, total sensitivity index of 25%), ocean vertical diffusivity ($\kappa_D$, 6%), and the interaction between the two (second-order interaction index of 3%). The simulated Antarctic ice sheet dynamics drive variance in flood risk via ice sheet bed slope before loading ("slope", 37%), equilibrium runoff line height ($h_0$, 29%), runoff line height sensitivity to temperature ($c$, 27%), mean annual Antarctic precipitation ($P_0$, 9%), and a proportionality constant for ice flow speed at the ice sheet grounding line ($f_0$, 6%). We also find substantial interactions among the Antarctic ice sheet parameters, ranging from 2-3% of the total flood risk variance.

**4 Discussion and caveats**

This analysis is intended as a didactic proof of concept and is silent on many likely important uncertainties. For example, we consider only a simple emulator of the fast Antarctic ice sheet dynamical contributions to sea level (Equation 1). Avenues to improve this analysis include to consider alternative model structure for this mechanism, including potentially a time-lagged response after the disintegration temperature has been reached. Our simple emulator also does not resolve contributions from the West Antarctic versus the East Antarctic ice sheets. This could be done by running two separately calibrated instances of the Antarctic ice sheet model employed here, and altering the assumed geometry, but is beyond the scope of the present work. We focus here on the potential impacts of the general Antarctic fast dynamics. Figure 2b demonstrates that this mechanism drives noticeable risks for New Orleans over the next 50 years, but Figure 4 highlights the fact that these AIS-driven risks are mild compared to those posed by storm surges.

The extrapolation of 36 years of tide gauge data to infer risks at the 1/100 or 1/500 level carries large uncertainties [e.g., *NOAA*, 2017]. We discuss further the potential implications of this limited supply of data in the Supplementary Text, where we describe the following sensitivity experiment. We estimated the distributions of 100-year surge height using 37-year subsets of tide gauge data from Galveston, Texas and Pensacola, Florida (the closest longer duration tide gauge locations to New Orleans). We found that the results obtained from the Grand Isle record are qualitatively similar to the results we might expect from other similar length data record from the U.S. Gulf Coast region. This ad hoc characterization of spatiotemporal representation uncertainty is in line with previous work which finds strong heterogeneity in the spatial and temporal distributions of surge height estimates along the Gulf Coast [e.g., *Buchanan et al.*, 2015]. A wide and realistic representation of uncertainty is needed, however, to provide low-confidence (and more specifically, not overconfident) future projections to inform decision-making [cf. *Herman et al.*, 2015; *Abadie et al.*, 2017]. Indeed, across all of our scenarios, only 32-39% of the SOW achieve the 500-year flood protection suggested for critical infrastructure, and about 72-79% of the SOW achieve the 100-year flood protection standard (c.f. Figure 3).

The results of the Sobol' sensitivity analysis indicate that the uncertainty in Antarctic ice sheet dynamics and statistical characterization of the storm surge are key drivers of the flood risk variability. The design of effective flood risk management strategies (i.e., the exogenous levee heightening "build" term) would benefit from tighter constraint on these geophysical mechanisms. This may come in the form of assimilating expert judgment regarding future Antarctic mass loss [e.g., *Oppenheimer et al.*, 2016] and consideration of alternative statistical approaches to characterize the storm surge level, for example, by leveraging new statistical models and improving the use of the limited data [e.g., *Naveau et al.*, 2016; *Stein*, 2017].

The sensitivity analysis demonstrates that over the 2015-2065 project period considered here, the RCP scenario and rapid Antarctic ice sheet dynamics (the parameters $T_{crit}$ and $\lambda$) are not strongly linked to variance in future flood risk (Figure 4). This is attributed to the fact that the RCP scenarios do not substantially diverge by mid-century [*Meinshausen et al.*, 2011]. If one considers a time horizon of 2100, for example, then RCP scenario and fast Antarctic mass loss likely play much larger roles. However, such a long time horizon requires consideration of dynamic adaptive strategies for risk management in light of the deep uncertainties characterized here [*Haasnoot et al.*, 2013; *Quinn et al.*, 2017], as opposed to designing a long-term strategy at the beginning of the study period. This is another important avenue for future research.

Optimal levee designs indicate that the relative contributions of overtopping versus piping or slope instability failures to the total levee failure probability vary depending on site-specific properties, but overtopping is the dominant failure mode [*Johnson et al.*, 2013; *Bischiniotis et al.*, 2016]. Our analysis considers overtopping to account for 100% of the total probability of failure, and serves to provide an upper bound on the system reliability (Figure ). We confirmed that our results are not significantly altered by prescribing overtopping to account for 60 or 80% of the total probability of failure, which is representative of the plausible range of failure modes [*Bischiniotis et al.*, 2016] (see Supporting Information Figures S2 and S3). A more detailed probabilistic treatment of the different failure modes outlined here, as well as pumping system or floodgate failure, is beyond the scope of this study. As a simplification, we use a single elevation, sea level, storm surge level, subsidence rate and levee height for the entire levee polder protecting the central metropolitan New Orleans area, and do not consider non-levee forms of flood protection (e.g., elevated structures). More detailed modeling would be appropriate to inform on-the-ground decision-making, accounting for local bathymetry, non-structural protection measures and levee system heterogeneity [e.g., *Fischbach et al.*, 2012; *Johnson et al.*, 2013]. These caveats point to important research needs, as well as why these results should not be used to directly inform on-the-ground decisions.

## 4 Conclusions

Given these caveats, our results indicate a level of flood risk in New Orleans that within 50 years may well exceed the 500-year flood protection target for critical infrastructure. This protection target is missed by the levee system alone across all scenarios (in the ensemble median). The ensemble median in all scenarios achieves the 100-year flood protection standard, but about 25% of the SOW miss the 100-year flood protection standard (Figure 3). The storm surge characterization is found to be the primary driver of variance in future flood risk, followed by Antarctic ice sheet dynamics (Figure 4). These results highlight the importance of investing resources to determine the appropriate level of protection for a given levee system, as well as the need for further research on statistical models for extreme storm surge levels. Indeed, our results demonstrate how pursuing reduced emissions trajectories can lead to more successful flood risk management strategies by mitigating deeply uncertain risks across multiple system components.

**Acknowledgments**

We gratefully acknowledge Greg Garner, Robert Fuller, Kelsey Ruckert, Benjamin Lee, Vivek Srikrishnan, Alexander Bakker, Rob Lempert, David Johnson, Neil Berg, as well as Bella and Chris Forest for invaluable inputs. This work was partially supported by the National Science

Foundation through the Network for Sustainable Climate Risk Management (SCRiM) under NSF cooperative agreement GEO-1240507 as well as the Penn State Center for Climate Risk Management. The authors are not aware of any real or perceived conflicts of interest. Any conclusions or recommendations expressed in this material are those of the authors and do not necessarily reflect the views of the funding agencies. Any errors and opinions are, of course, those of the authors. All model codes, analysis codes, data and model output used for analysis are freely available from https://github.com/scrim-network/BRICK/tree/scenarios.

**Author Contributions**

TW and KK initiated the study and designed the research. TW produced the model simulations, designed the initial figures and wrote the first draft. Both contributed to the final text and figures.

Tables

| Radiative forcing | Antarctic fast dynamics | Storm surge |
|---|---|---|
| RCP2.6 | None | Stationary |
| RCP4.5 | Uniform prior pdf | Non-stationary |
| RCP8.5 | Gamma prior pdf | |

**Table 1.** Deeply uncertain factors sampled to construct scenarios.

# Figures

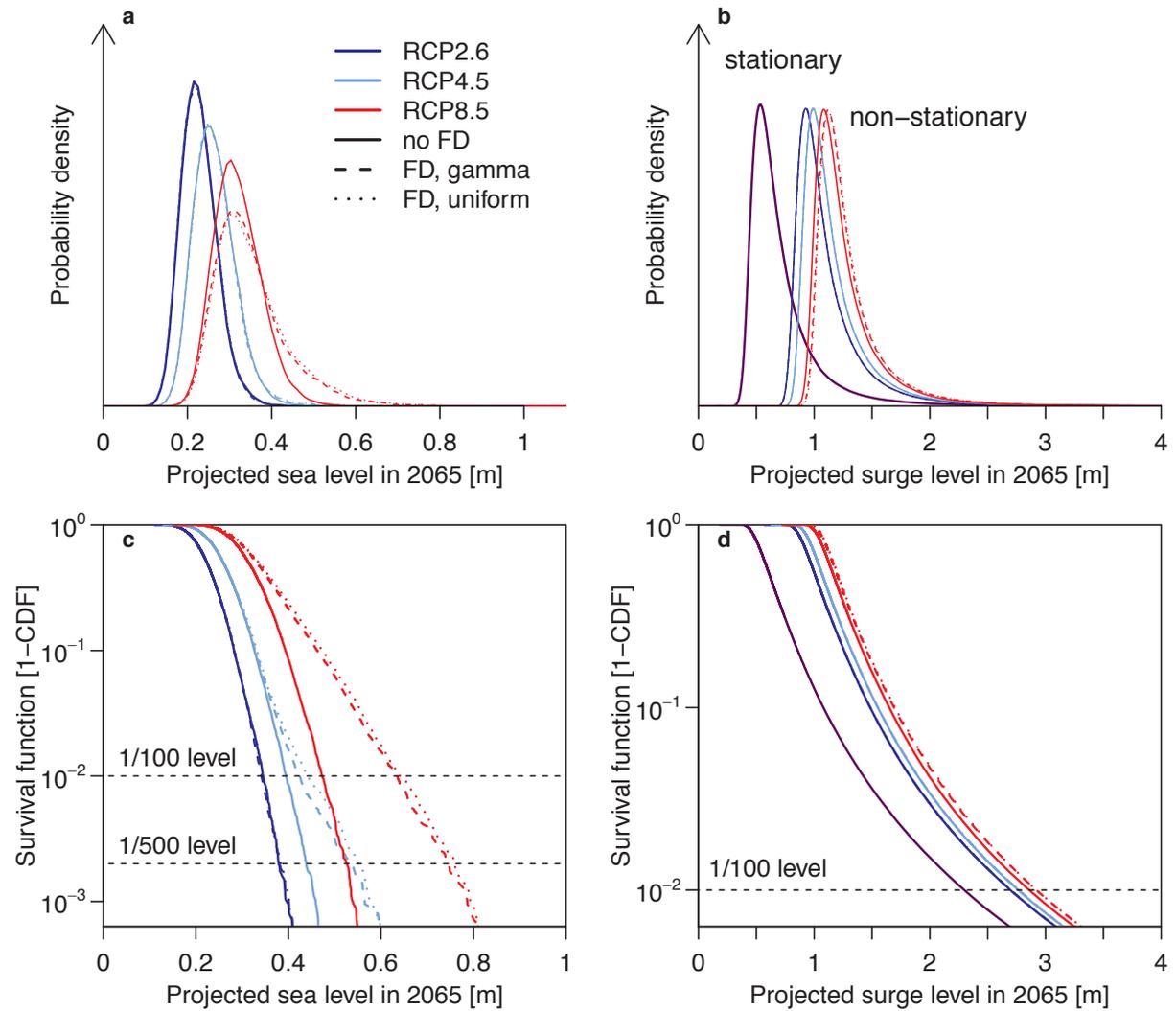

**Figure 1.** Probability density functions of (a) projected future sea level and (b) projected future storm surge levels, and (c, d) their associated survival functions (which give the total probability in the right tail of a distribution). "FD" refers to the fast Antarctic ice sheet dynamics; the dashed and dotted curves refer to the distributions obtained using the gamma and uniform prior distributions (respectively) for the fast dynamics parameters.

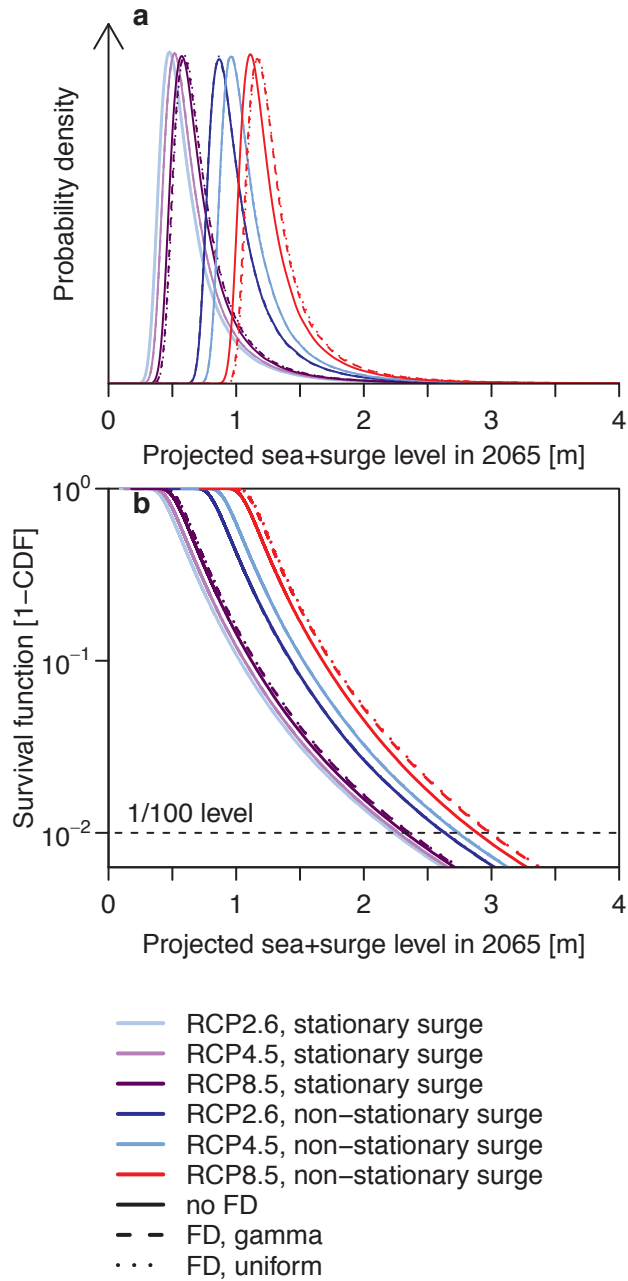

**Figure 2.** (a) Probability density functions and (b) survival functions for the 18 scenarios of radiative forcing, Antarctic ice sheet fast dynamics and storm surge characterization. "FD" refers to the fast Antarctic ice sheet dynamics; the dashed and dotted curves refer to the distributions obtained using the gamma and uniform prior distributions (respectively) for the fast dynamics parameters.

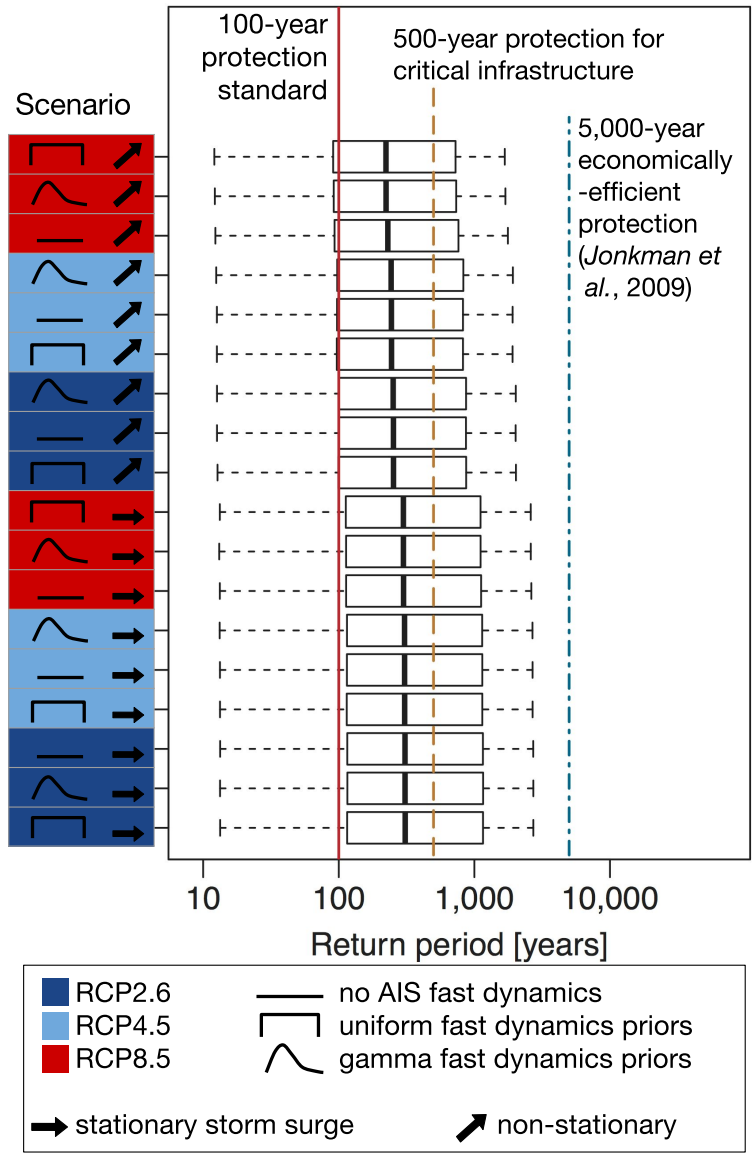

**Figure 3.** Return periods (inverse of flood probability) for the 18 scenarios, ordered from most severe (top) to least severe (bottom). The bold lines within each box denote the ensemble median; the extent of the boxes denotes the interquartile range (25-75%); the extent of the horizontal dashed lines denotes the ensemble extrema; the solid red line denotes the 100-year protection standard; the dashed yellow line denotes the 500-year protection level for critical infrastructure [Coastal Protection and Restoration Authority of Louisiana, 2017]; and the blue dot-dashed line denotes the economically-efficient protection level [Jonkman et al., 2009].

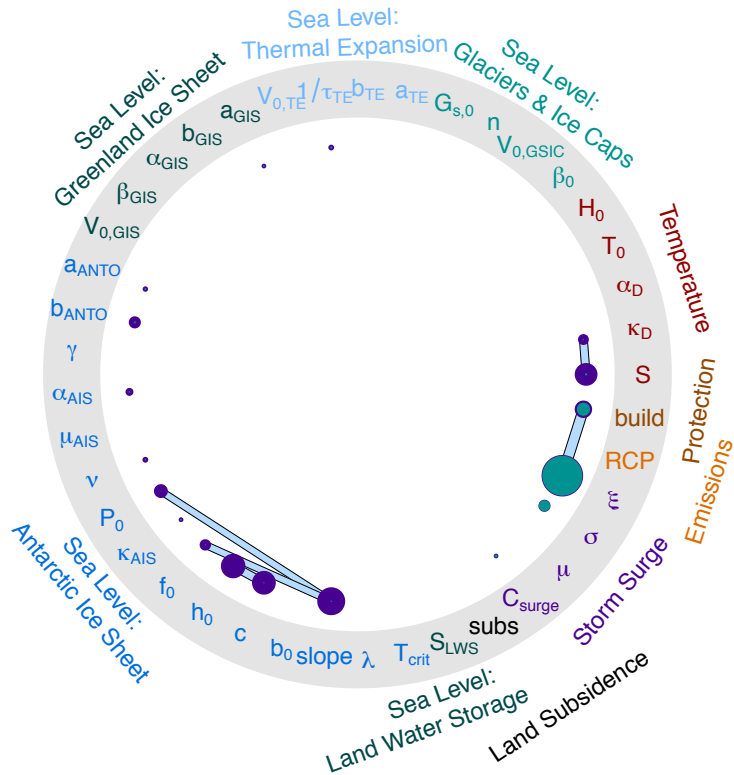

**Figure 4.** Sobol' sensitivity results for decomposition of the variance in projected flood risk over the 2015-2065 period (mean annual exceedance probability). Filled blue nodes represent first-order sensitivity indices; filled purple nodes represent total-order sensitivity indices; filled gray bars represent second-order sensitivity indices for the interaction between the parameter pair.

Supporting Information for

**Deep Uncertainty Surrounding Coastal Flood Risk Projections: A Case Study for New Orleans**

Tony E. Wong[1] and Klaus Keller[1,2,3]

[1]Earth and Environmental Systems Institute, Pennsylvania State University, University Park, PA 16802, USA.

[2]Department of Geosciences, Pennsylvania State University, University Park, PA 16802, USA.

[3]Department of Engineering and Public Policy, Carnegie Mellon University, Pittsburgh, PA 15289, USA.

**Contents of this file**

**Introduction**

This supporting information provides further details regarding the statistical modeling of the distribution of storm surge levels, as well as a series of figures to support the main conclusions of the article. The supplemental table provides a summary of the endogenous parameters and exogenous forcings considered in the Sobol' sensitivity analysis (Figure 4, main text). The supplemental figures include an evaluation of the impacts of the planned addition of 0.91 meters (3 feet) of dike heightening on the projected flood risk (Figure S1), and we illustrate the reduction in projected return period (i.e., decrease in reliability) caused by non-overtopping modes of dike failure (Figures S2 and S3). We conduct two supplemental sensitivity analyses to assess the extent to which large uncertainties in the Antarctic ice sheet dynamics (runoff/ablation line height, Figure S4) and the storm surge GEV parameter estimates (Figure S5) impact the decomposition of flood risk variance. Finally, we conduct sensitivity experiments assessing the extent to which spatio-temporal representation uncertainty may affect return level estimates using as short a data record as is available at Grand Isle, Louisiana (Figures S6 and S7).

**Text S1.**

Here, we provide additional details regarding the maximum likelihood estimates (MLE) and Markov chain Monte Carlo (MCMC) estimates of the Generalized Extreme Value (GEV) distribution parameters fit to the tide gauge data from Grand Isle, Louisiana [*NOAA*, 2017]. We also discuss the spatio-temporal representation uncertainty in the tide gauge data from the United State Gulf Coast.

**Statistical modeling**

First, we subtract annual means from the tide gauge record. Second, we calculate annual block maxima from the detrended record. The MLE method calculates the likelihood of an assumed set of GEV parameters as the joint probability of observing the detrended set of annual block maxima, given the assumed GEV parameters. The quasi-Newton, gradient-based "BFGS" (Broyden–Fletcher–Goldfarb–Shanno) optimization algorithm is used to iterate until convergence to maximal likelihood is achieved within a relative tolerance of $10^{-8}$ [*Broyden*, 1970]. By sampling from the likelihood surface to be optimized, we obtain also estimates of the 5-95% confidence intervals for each GEV parameter. These 5-95% confidence intervals are assumed to represent (roughly) a +/-$2\sigma$ window around the central estimate of each parameter. Dividing the 5-95% confidence intervals by 4 yields estimates of the standard deviation for each GEV parameter. These are 13.5 mm, 0.2 log(mm), and 0.25 for the $\mu$ (location), log($\sigma$) (scale) and $\xi$ (shape) GEV parameters. In order to maintain $\sigma$ >0, we estimate log($\sigma$) during the MCMC simulation.

We use these estimates as the starting point for the standard deviations for the normally-distributed (with mean zero) step-sizes for the Gaussian random walk MCMC simulation of the GEV parameters. We tuned the standard deviation associated with proposals for $\mu$ to 22 mm and keep the standard deviations for proposals for log($\sigma$) and $\xi$ at 0.15 log(mm) and 0.23, respectively. These proposal standard deviations achieve acceptance rates for each parameter around 0.44, which is approximately optimal [*Rosenthal*, 2011] for the type of "Metropolis-within-Gibbs" sampling approach implemented here [*Gilleland et al.*, 2013]. We use the maximum likelihood estimates as initial parameter estimates for the MCMC simulation. Preliminary experiments using dispersed initial conditions indicated that our results are robust to starting points (results not shown). We adopt prior distributions for the GEV location ($\mu$) and scale ($\sigma$) parameters as normal with mean equal to the MLE parameters found above, and standard deviation equal to four times the standard deviation estimates. We use a uniform prior distribution between 0 and 5 for the shape parameter ($\xi$). These priors strike a balance between confidence in the maximum likelihood GEV parameter estimates and exploring the parameter-space. We use the uniform prior for the shape parameter to disallow negative values, wherein the fundamental behavior of the GEV distribution changes.

This algorithm samples each GEV parameter successively for 500,000 iterations, for two independent Markov chain Monte Carlo simulations. The first 100,000 iterations are removed for burn-in. We randomly draw 12,586 samples without replacement from the remaining combined 800,000 samples to serve as the ensemble for analysis. We use visual inspection, dispersed initial values, Gelman and Rubin diagnostics, and Heidelberger and Welch diagnostics to assess convergence [*Heidelberger et al.*, 1981; *Heidelberger and Welch*, 1983; *Gelman and Rubin*, 1992].

**Spatio-temporal representation uncertainty**

We conduct an experiment to evaluate the sensitivity of our results to limited available data. We use the tide gauge records from Galveston, TX (101 years) and Pensacola, FL (83 years), which are the closest other longer duration tide gauge sites [*Caldwell et al.*, 2015]. Both have relatively low amounts of missing data; we use only years with at least 90% of the data present in our analysis. We break each tide gauge record up into blocks of 37 years. The blocks are offset by 8 years from one another so that adjacent blocks contain about 75% overlap with one another. Thus, there are nine blocks for Galveston and seven blocks for Pensacola. We obtain calibrated GEV parameters using MCMC (similarly to the main experiments in the study) and draw 12,586 samples for analysis from the posterior distributions. This sample size was selected for consistency with the resolution of the ensembles presented in the main experiments for New Orleans. For each ensemble member at each of the two sites, within each block experiment, we calculate the surge height associated with the 100-year return period (i.e., the 100-year flood height). This results in a probability density function (pdf) for each block experiment, for each site. We also calculate the pdf of 100-year return levels for the 36 years of tide gauge data from Grand Isle (New Orleans), Louisiana.

Supplementary Figure S6 (below) shows a subset of these distributions for Galveston (a) and for Pensacola (b), with the distribution from New Orleans superimposed (dashed black pdf). We find substantial temporal variability in the estimates of the 100-year flood level at Galveston, but much less variability at Pensacola (Supplementary Figure S7). These sensitivity results motivate the need to include non-stationarity in modeling approaches for extreme sea levels. The New Orleans pdf fits well within the range of pdfs fit from 37-year records in the Gulf Coast region, at the low end of estimates from Galveston (to the west of New Orleans) and at the high end of estimates from Pensacola (to the east). This sensitivity experiment suggests that, while there is substantial spatio-temporal representation uncertainty, the data record from Grand Isle is qualitatively similar to the results we might expect from any other record of similar length from this region.

**Table S1.** Endogenous model parameters and exogenous forcings considered in the Sobol' sensitivity analysis.

| Parameter name | Parameter description |
| --- | --- |
| $S$ | Equilibrium climate sensitivity to doubling $CO_2$ (°C) |
| $\kappa_D$ | Ocean vertical diffusivity (cm$^2$ s$^{-1}$) |
| $\alpha_D$ | Aerosol radiative forcing scaling factor (-) |
| $T_0$ | Initial temperature anomaly (°C) |
| $H_0$ | Initial ocean heat uptake ($10^{22}$ J) |
| $\beta_0$ | Initial glacier/ice cap mass balance temperature sensitivity (m y$^{-1}$ °C$^{-1}$) |
| $V_{0,GSIC}$ | Initial glacier/ice cap volume (m sea level equivalent (SLE)) |
| $n$ | Exponent for glacier/ice cap area-volume scaling (-) |
| $G_{s,0}$ | Initial condition for glacier/ice cap sea-level contribution (m SLE) |
| $a_{TE}$ | Equilibrium thermal expansion temperature sensitivity (m °C$^{-1}$) |
| $b_{TE}$ | Equilibrium thermal expansion for 0 °C temperature anomaly (m SLE) |
| $1/\tau_{TE}$ | Inverse timescale for thermal expansion response (y$^{-1}$) |
| $V_{0,TE}$ | Initial condition for thermal expansion contribution to sea level (m SLE) |
| $a_{GIS}$ | Equilibrium Greenland ice sheet volume temperature sensitivity (m °C$^{-1}$) |
| $b_{GIS}$ | Equilibrium Greenland ice sheet volume (m SLE) |
| $\alpha_{GIS}$ | Greenland ice sheet response timescale temperature sensitivity (°C$^{-1}$ y$^{-1}$) |
| $\beta_{GIS}$ | Greenland ice sheet response timescale temperature sensitivity (y$^{-1}$) |
| $V_{0,GIS}$ | Initial condition for Greenland ice sheet sea level contribution (m SLE) |
| $a_{ANTO}$ | Antarctic ocean temperature sensitivity to global temperature (°C °C$^{-1}$) |
| $b_{ANTO}$ | Equilibrium Antarctic ocean temperature (°C) |
| $\gamma$ | Power for relation of Antarctic ice flow speed to water depth (-) |
| $\alpha_{AIS}$ | Effect of ocean subsurface temperature on ice flux partition parameter (-) |
| $\mu_{AIS}$ | Parabolic ice surface profile parameter (m$^{0.5}$) |
| $\nu$ | Antarctic runoff and precipitation proportionality constant (m$^{-0.5}$ y$^{-0.5}$) |
| $P_0$ | Annual Antarctic precipitation for surface temperature 0 °C (m) |
| $\kappa_{AIS}$ | Coefficient, dependency of Antarctic precipitation on temperature (°C$^{-1}$) |
| $f_0$ | Antarctic ice flow at grounding line proportionality constant (m y$^{-1}$) |
| $h_0$ | Antarctic runoff line height at 0 °C surface temperature (m) |
| $c$ | Antarctic runoff line height temperature sensitivity (m °C$^{-1}$) |
| $b_0$ | Undisturbed bed height at Antarctic continent center (m) |
| $slope$ | Slope of Antarctic ice sheet bed before ice loading (-) |
| $\lambda$ | Fast Antarctic dynamic disintegration rate (m) |
| $T_{crit}$ | Fast Antarctic dynamic disintegration trigger temperature (°C) |
| $S_{LWS}$ | Mean annual land water storage contribution to sea level (m) |
| $subs$ | Mean annual land subsidence rate (mm y$^{-1}$) |
| $C_{surge}$ | Proportionality factor relating increases in storm surge and sea levels (-) |
| $\mu$ | Location, storm surge generalized extreme value distribution (mm) |
| $\sigma$ | Location, storm surge generalized extreme value distribution (mm) |
| $\xi$ | Location, storm surge generalized extreme value distribution (-) |
| RCP | Representative Concentration Pathway forcing (2.5, 4.5, 6.0, or 8.5) |
| $build$ | Additional heightening of the levee system (m) |

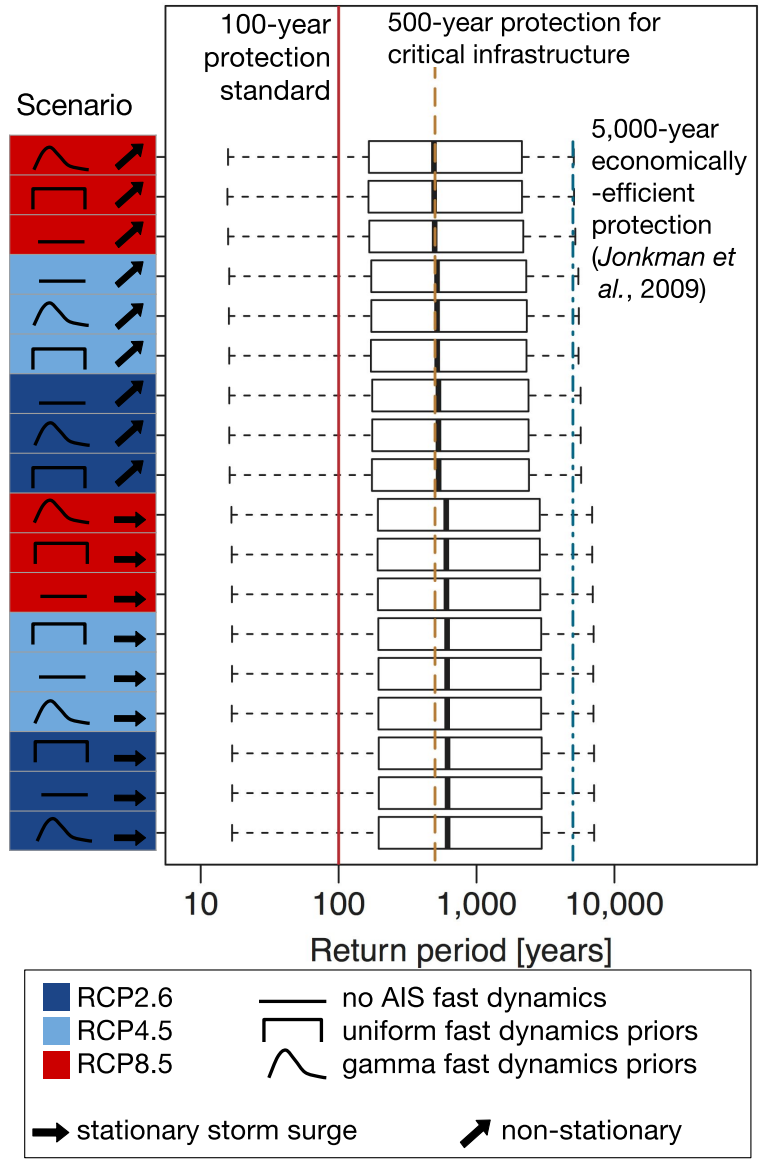

**Figure S1.** Return periods (inverse of flood probability) for the 18 scenarios, ordered from most severe (top) to least severe (bottom), with an additional 0.91 meters (3 feet) of dike heightening. The bold lines within each box denote the ensemble median; the extent of the boxes denotes the interquartile range (25-75%); the extent of the horizontal dashed lines denotes the ensemble extrema; the solid red line denotes the 100-year protection standard; the dashed yellow line denotes the 500-year protection level for critical infrastructure [*Coastal Protection and Restoration Authority of Louisiana*, 2017]; and the blue dot-dashed line denotes the economically-efficient protection level [*Jonkman et al.*, 2009].

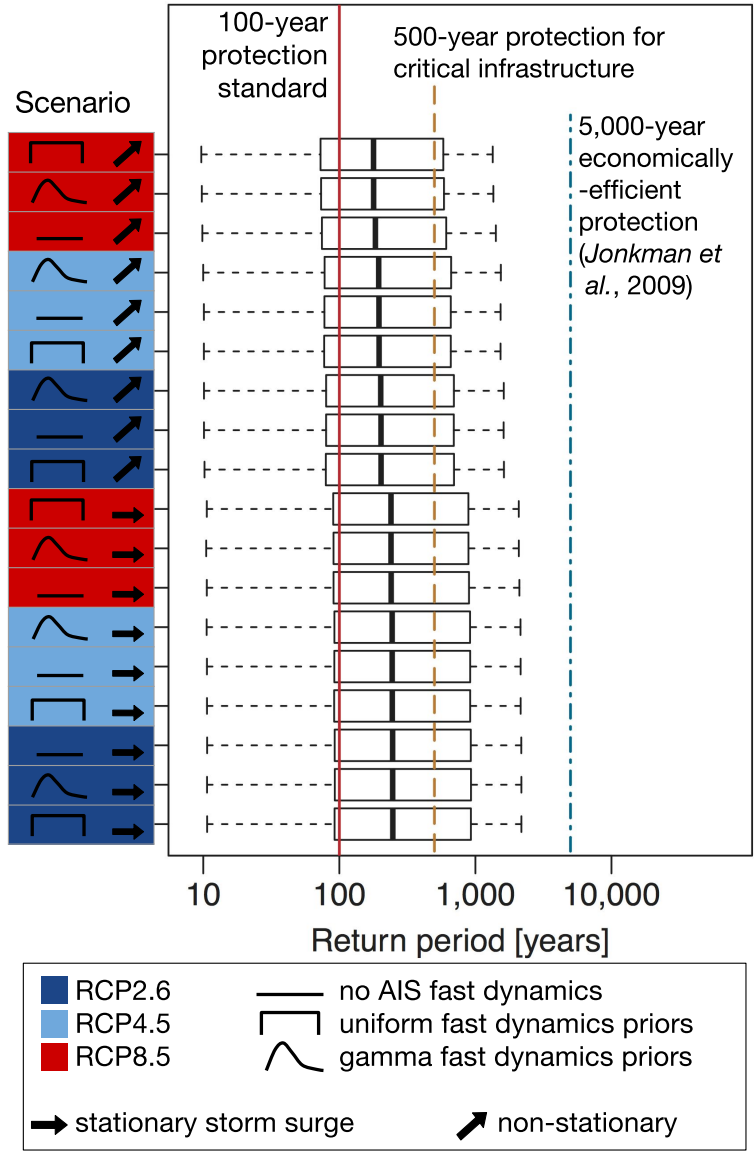

**Figure S2.** Return periods (inverse of flood probability) for the 18 scenarios, ordered from most severe (top) to least severe (bottom), with overtopping accounting for 80% of total failure probability. The bold lines within each box denote the ensemble median; the extent of the boxes denotes the interquartile range (25-75%); the extent of the horizontal dashed lines denotes the ensemble extrema; the solid red line denotes the 100-year protection standard; the dashed yellow line denotes the 500-year protection level for critical infrastructure [*Coastal Protection and Restoration Authority of Louisiana*, 2017]; and the blue dot-dashed line denotes the economically-efficient protection level [*Jonkman et al.*, 2009].

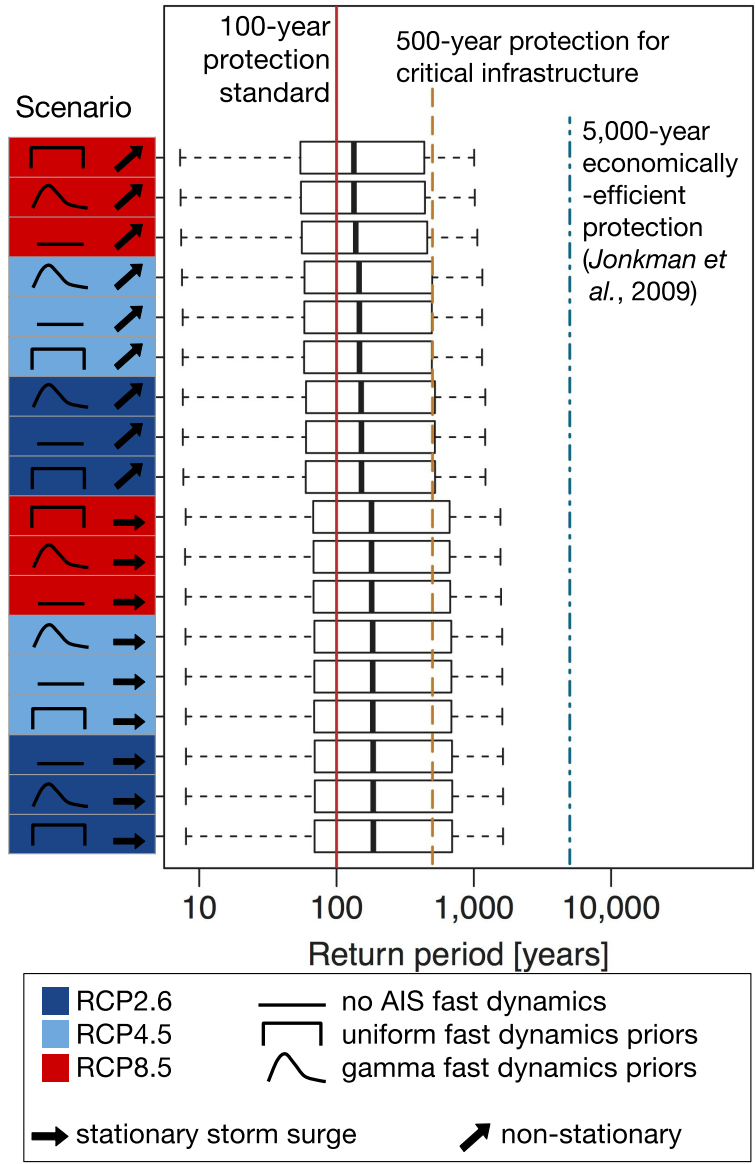

**Figure S3.** Return periods (inverse of flood probability) for the 18 scenarios, ordered from most severe (top) to least severe (bottom), with overtopping accounting for 60% of total failure probability. The bold lines within each box denote the ensemble median; the extent of the boxes denotes the interquartile range (25-75%); the extent of the horizontal dashed lines denotes the ensemble extrema; the solid red line denotes the 100-year protection standard; the dashed yellow line denotes the 500-year protection level for critical infrastructure [*Coastal Protection and Restoration Authority of Louisiana*, 2017]; and the blue dot-dashed line denotes the economically-efficient protection level [*Jonkman et al.*, 2009].

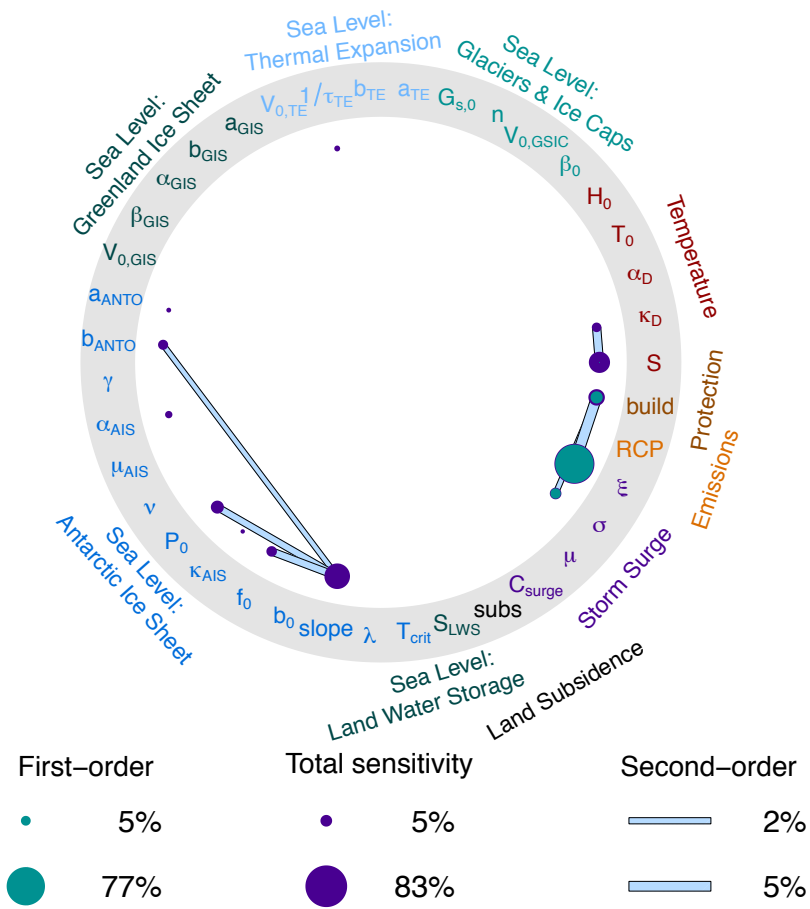

**Figure S4.** Sobol' sensitivity results for decomposition of the variance in projected flood risk over the 2015-2065 period (mean annual exceedance probability), with the Antarctic ice sheet runoff line height parameters ($h_0$ and $c$) omitted from the analysis. Filled blue nodes represent first-order sensitivity indices; filled purple nodes represent total-order sensitivity indices; filled gray bars represent second-order sensitivity indices for the interaction between the parameter pair.

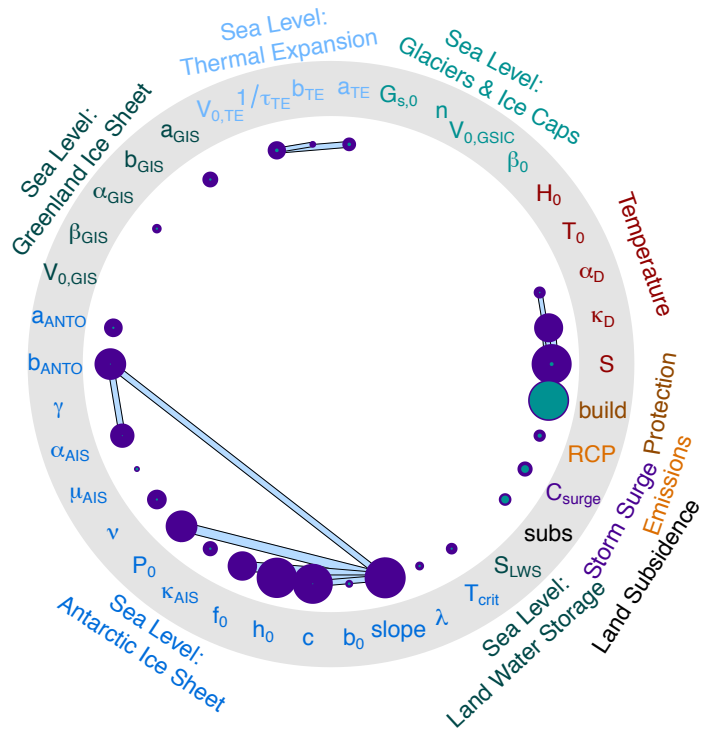

**Figure S5.** Sobol' sensitivity results for decomposition of the variance in projected flood risk over the 2015-2065 period (mean annual exceedance probability), with the storm surge generalized extreme value distribution parameters ($\mu$, $\sigma$, and $\xi$) omitted from the analysis. Filled blue nodes represent first-order sensitivity indices; filled purple nodes represent total-order sensitivity indices; filled gray bars represent second-order sensitivity indices for the interaction between the parameter pair.

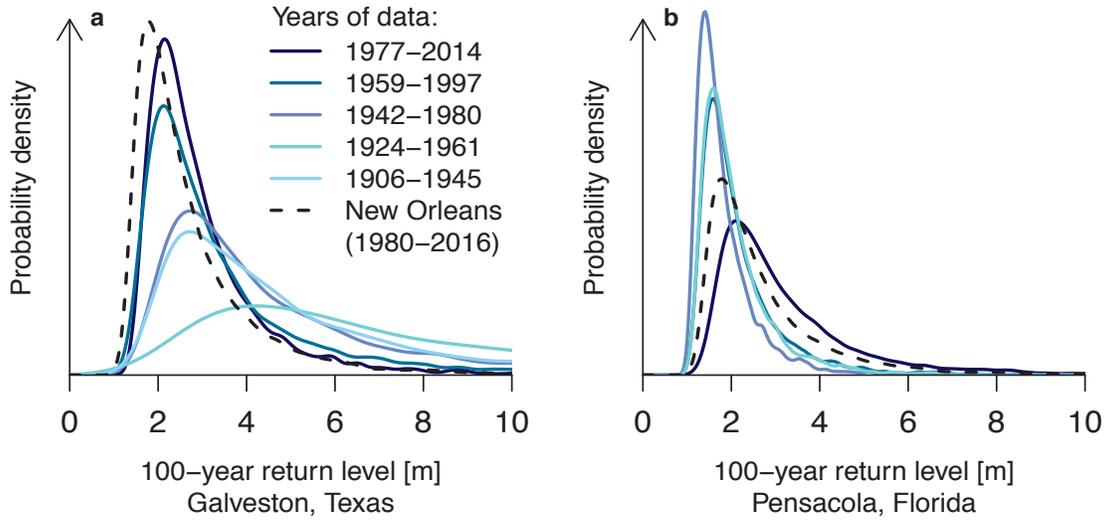

**Figure S6.** Distributions of estimated 100-year return levels for (a) Galveston, Texas and (b) Pensacola, Florida, using 37-year blocks of tide gauge data for the calibration. Comparison of the multiple distributions within each panel characterizes the temporal representation uncertainty, while comparison of the distributions between the two panels, as well as with the New Orleans data (dashed curve, superimposed) characterizes the spatial representation uncertainty.

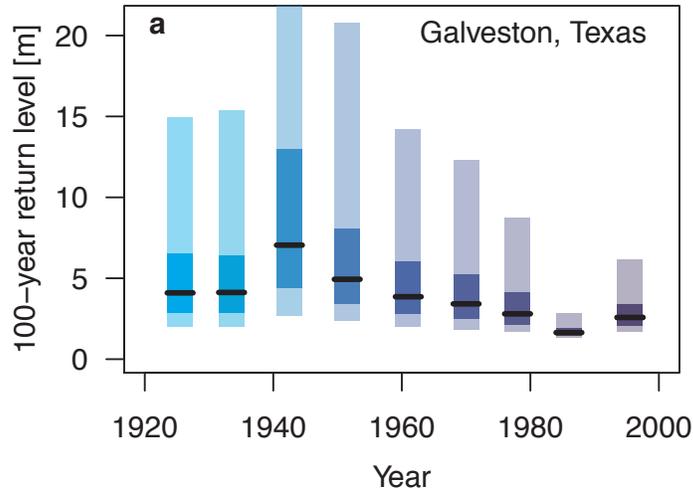

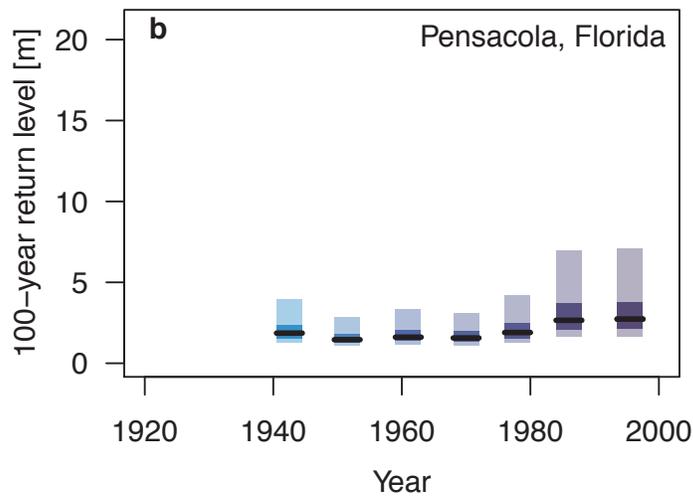

**Figure S7.** Distributions of estimated 100-year return levels for (a) Galveston, Texas and (b) Pensacola, Florida, using 37-year blocks of tide gauge data for the calibration. Each vertical box represents a distribution, centered at the central year from that 37-year subset of tide gauge data used for the calibration. The black lines give the ensemble median, the darker shaded box gives the 25-75% credible range and the lighter shaded box gives the 5-95% range.